\documentclass[aps,prl,twocolumn,showpacs,superscriptaddress,amsmath,amssymb,longbibliography]{revtex4-1}

\usepackage{amsmath}
\usepackage{amssymb}
\usepackage{graphicx}
\usepackage{float}
\usepackage{color}

\usepackage[final]{pdfpages}

\makeatletter
\AtBeginDocument{\let\LS@rot\@undefined}
\makeatother

\begin{document}

\title{Dynamics of a network fluid within the liquid-gas coexistence region}

  \author{C. S. Dias}
   \email{csdias@fc.ul.pt}
    \affiliation{Departamento de F\'{\i}sica, Faculdade de Ci\^{e}ncias, Universidade de Lisboa, 
    1749-016 Lisboa, Portugal}
    \affiliation{Centro de F\'{i}sica Te\'{o}rica e Computacional, Universidade de Lisboa, 
    1749-016 Lisboa, Portugal}

  \author{J. M. Tavares}
   \email{jmtavares@fc.ul.pt}
    \affiliation{Centro de F\'{i}sica Te\'{o}rica e Computacional, Universidade de Lisboa, 
    1749-016 Lisboa, Portugal}
    \affiliation{Instituto Superior de Engenharia de Lisboa, ISEL, Avenida Conselheiro Em\'{i}dio Navarro, 1  1950-062 Lisboa, Portugal}
    
  \author{N. A. M. Ara\'ujo}
   \email{nmaraujo@fc.ul.pt}
    \affiliation{Departamento de F\'{\i}sica, Faculdade de Ci\^{e}ncias, Universidade de Lisboa, 
    1749-016 Lisboa, Portugal}
    \affiliation{Centro de F\'{i}sica Te\'{o}rica e Computacional, Universidade de Lisboa, 
    1749-016 Lisboa, Portugal}
    
  \author{M. M. Telo da Gama}
   \email{mmgama@fc.ul.pt}
    \affiliation{Departamento de F\'{\i}sica, Faculdade de Ci\^{e}ncias, Universidade de Lisboa, 
    1749-016 Lisboa, Portugal}
    \affiliation{Centro de F\'{i}sica Te\'{o}rica e Computacional, Universidade de Lisboa, 
    1749-016 Lisboa, Portugal}

\begin{abstract}
Low-density networks of molecules or colloids are formed at low temperatures
when the interparticle interactions are valence limited. Prototypical examples
are networks of patchy particles, where the limited valence results from highly
directional pairwise interactions. We combine extensive Langevin simulations
and Wertheim's theory of association to study these networks. We find a scale-free (relaxation)
dynamics within  the liquid-gas coexistence region, which differs from that usually observed for 
isotropic particles. While for isotropic particles the relaxation 
dynamics is driven by surface tension (coarsening), when the valence is limited, 
the slow relaxation proceeds through the formation of an intermediate non-equilibrium gel via a geometrical 
percolation transition in the Random Percolation universality class.

\end{abstract}

\maketitle

The formation of stable low-density structures of interconnected molecules or
colloids is an open question that attracted researchers in the last decade
\cite{Dobnikar2013,Zaccarelli2007,Elliott2003,Ruzicka2011,Sciortino2004}.
These unique structures are simultaneously of low packing fraction and
resilient to mechanical perturbations. The former can be achieved through
limited valence of the individual constituents. The latter requires
sufficiently strong interparticle bonding. However, establishing strong bonds
leads to rough energy landscapes, challenging the formation of the final
equilibrium structures \cite{Chakrabarti2014,Zaccarelli2006}.  Kinetically
arrested structures are usually obtained, which may be significantly different
from the thermodynamic ones.
\begin{figure}
   \begin{center} \includegraphics[width=0.9\columnwidth]{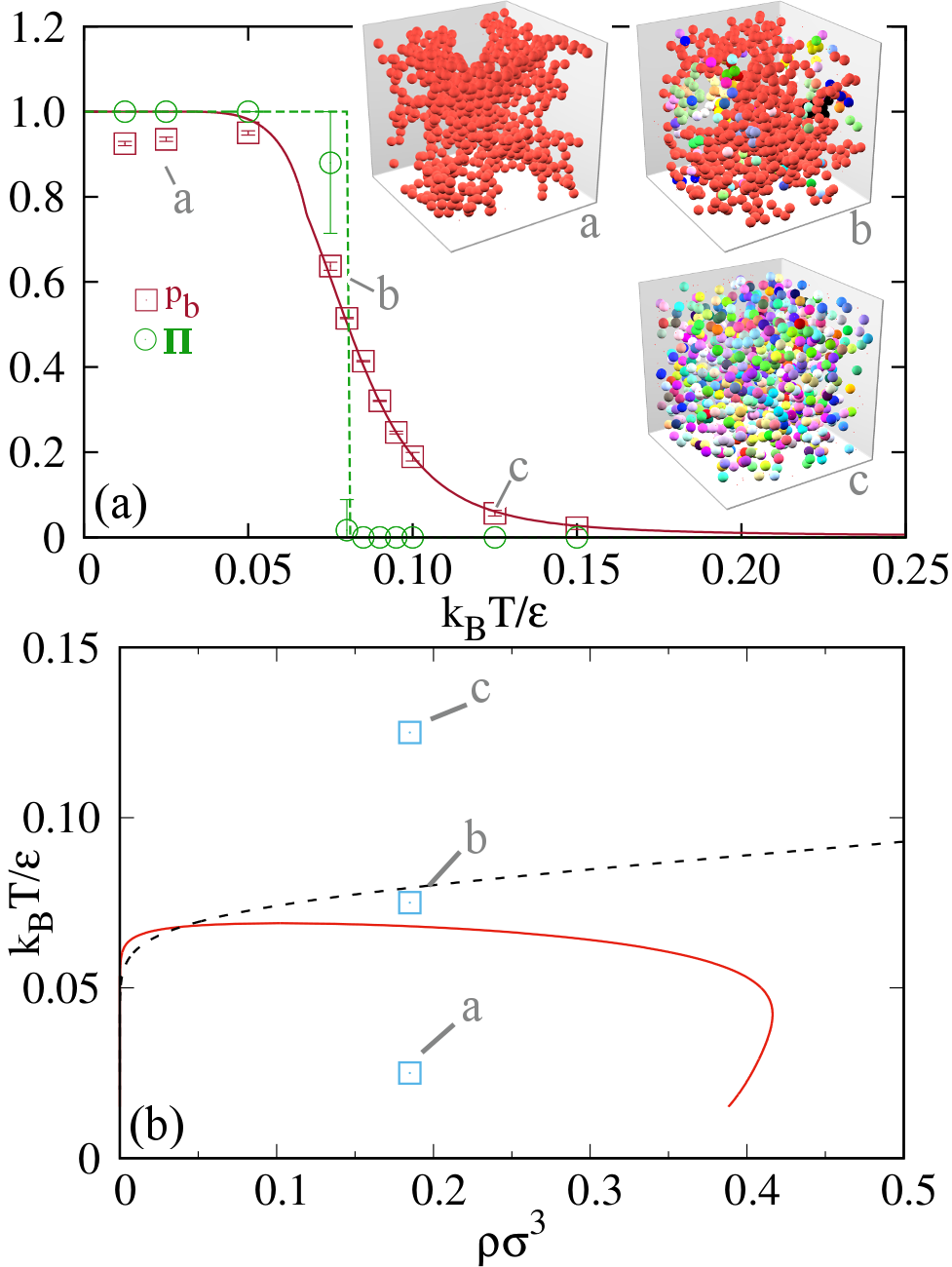} \\
\caption{(a) The (red) squares are the bonding fraction $p_b$ as a function of the
rescaled temperature $k_BT/\varepsilon$, obtained by extrapolating the time-dependence 
of $p_b$ to its asymptotic value from the numerical data for systems of linear size  $L=\{8,16,32\}$. 
The solid line is the equilibrium value
predicted by Wertheim's first-order perturbation theory of association (see
Methods for further details). The (green) circles are
the wrapping probability $\Pi$, defined numerically as the fraction of samples
with an aggregate of bonded particles that touches simultaneously the top and
the bottom boundaries of the simulation box. The dashed line corresponds to the
mean-field prediction for $\Pi$, which is a step function at $k_BT^*/\varepsilon$, where
$p_b(k_BT^*/\varepsilon)=1/2$. Simulations were performed at a density $0.2\sigma^{-3}$. 
(b) Density-temperature phase diagram, where the density $\rho=N/V$ is 
the number of particles per unit volume and the temperature is 
the rescaled temperature $k_BT/\varepsilon$. The solid (red) line is the 
coexistence line and the dashed (black) line is the percolation line.
Insets are snapshots at $k_BT/\varepsilon$f
a) 0.025, b) 0.075, and c) 0.125, where different colors correspond to
different aggregates of bonded particles. Numerical results were obtained by
averaging over $10$ samples, for a box of linear size $L=16$, in units of the
particle diameter, and $N=800$ particles.} 
\label{fig.diagram}
   \end{center}
\end{figure}

Colloidal valence may be engineered by chemically patterning the colloidal
surfaces with attractive patches
\cite{Zhang2014a,Paulson2015,Lu2013,Duguet2011,Hu2012,Kretzschmar2011,Sacanna2011,Solomon2011,Pawar2010,Sacanna2013a,Manoharan2015}.
Studies of equilibrium phase diagrams have shown that at low valence, due to the decrease in the density of the liquid binodal, 
a gel-like low density structure is thermodynamically stable (equilibrium) for a wide range of attractive strengths and densities \cite{Zaccarelli2005}. 
However, inside the coexistence region, the system is expected to phase separate into colloid-rich and colloidal-poor phases 
\cite{Mahynski2015,Noya2015,Romano2014,Russo2009,Bianchi2006,Bianchi2011}.
Nevertheless, as the lifetime of thermal reversible bonds increases
exponentially with the attraction, an arrested gel is formed
\cite{Kroy2004} in line with experimental observations for
mixtures of colloids and polymers \cite{Manley2005}. These non-ergodic gels
are significantly different from the thermodynamic ones.

\begin{figure}[t]
   \begin{center}
\includegraphics[width=0.9\columnwidth]{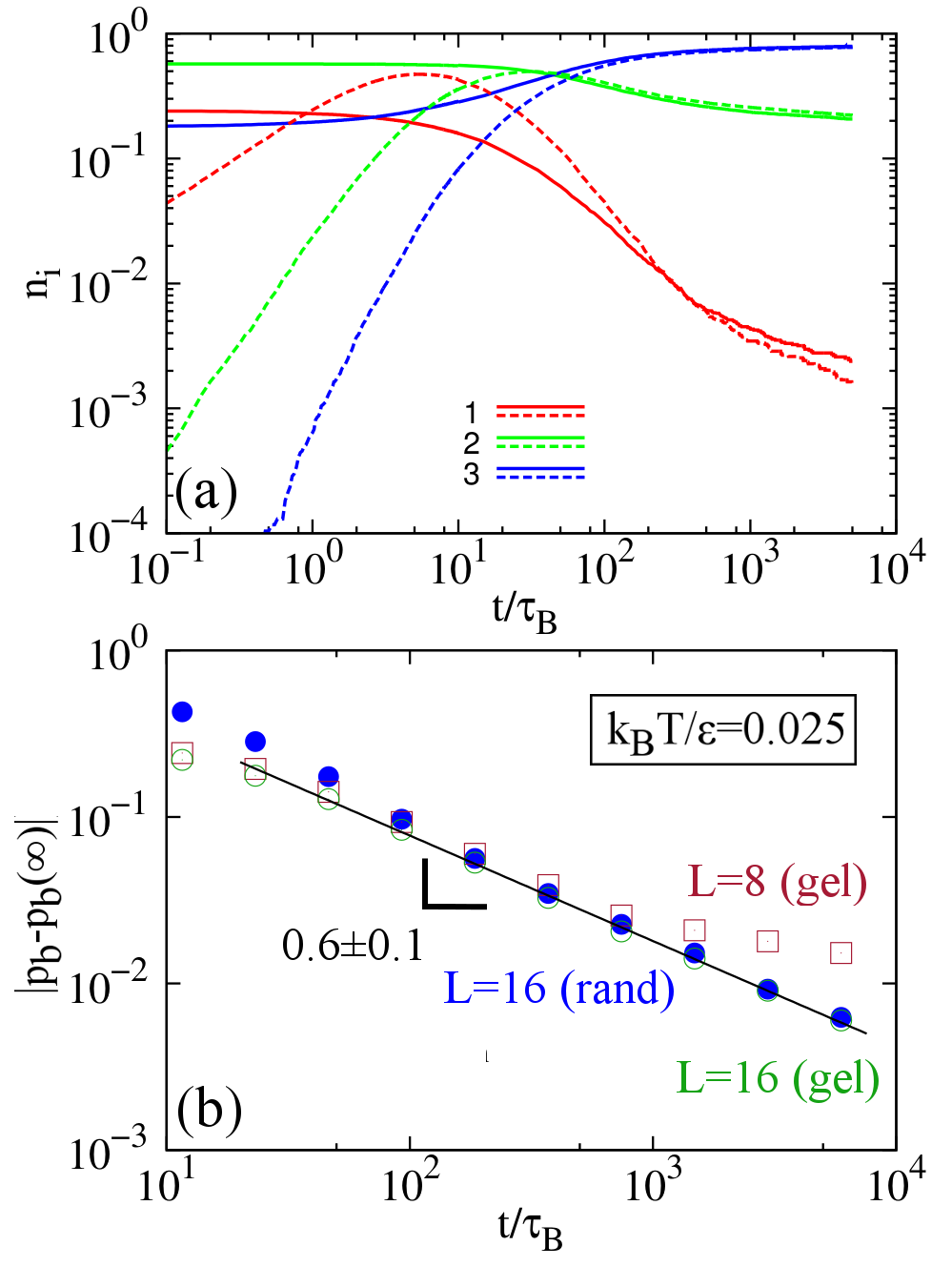} \\
\caption{(a) Time evolution of the fraction of particles with one (red), two (green), and three (blue) 
bonds, at $k_BT/\varepsilon=0.025$ (below the coexistence curve). The solid lines are for the relaxation of a tree-like gel grown by 
ballistic aggregation and loopless irreversible bond formation. The dashed lines are for the aggregation of an initial fully 
unbonded suspension of patchy particles. Time is rescaled by the Brownian time $\tau_B=\sigma^2/D_t$, where $\sigma$ 
is the particle diameter and $D_t$ is the self-diffusion coefficient of a single particle at short times. 
Results are averages over $100$ samples for $t/\tau_B<10$ and and $20$ samples for $t/\tau_B>10$, 
of a cubic box of length $L=16$, in units of the particle diameter. (b) Convergence of the bonding fraction to the 
asymptotic value $|p_b-p_b(\infty)|$ at low temperature ($k_BT/\varepsilon<0.075$), 
for initial conditions of tree-like gels grown by ballistic aggregation (gel) and fully 
unbonded suspensions (rand), where the time is rescaled by the Brownian time $\tau_B=\sigma^2/D_t$, 
obtained by averaging over $20$ samples in a box of length $L=\{8,16\}$ (same density), 
in units of the particle diameter. Simulations were performed at a density $0.2\sigma^{-3}$.} \label{fig.valence}
   \end{center}
\end{figure}
In what follows, we investigate the collective dynamics of limited-valence
particles in various regions of the phase diagram. We show that the interplay
between relaxation mechanisms at different length and time scales leads to
distinct dynamics at low and high temperatures. At high temperatures the
dynamics is driven by sequences of bond breaking and forming events and it is
characterized by an Arrhenius relaxation time related to the thermal
reversibility of the bonds. At low temperatures, however, a scale-free dynamics is observed. 
  
\begin{figure*}[t]
   \begin{center} \includegraphics[width=0.8\textwidth]{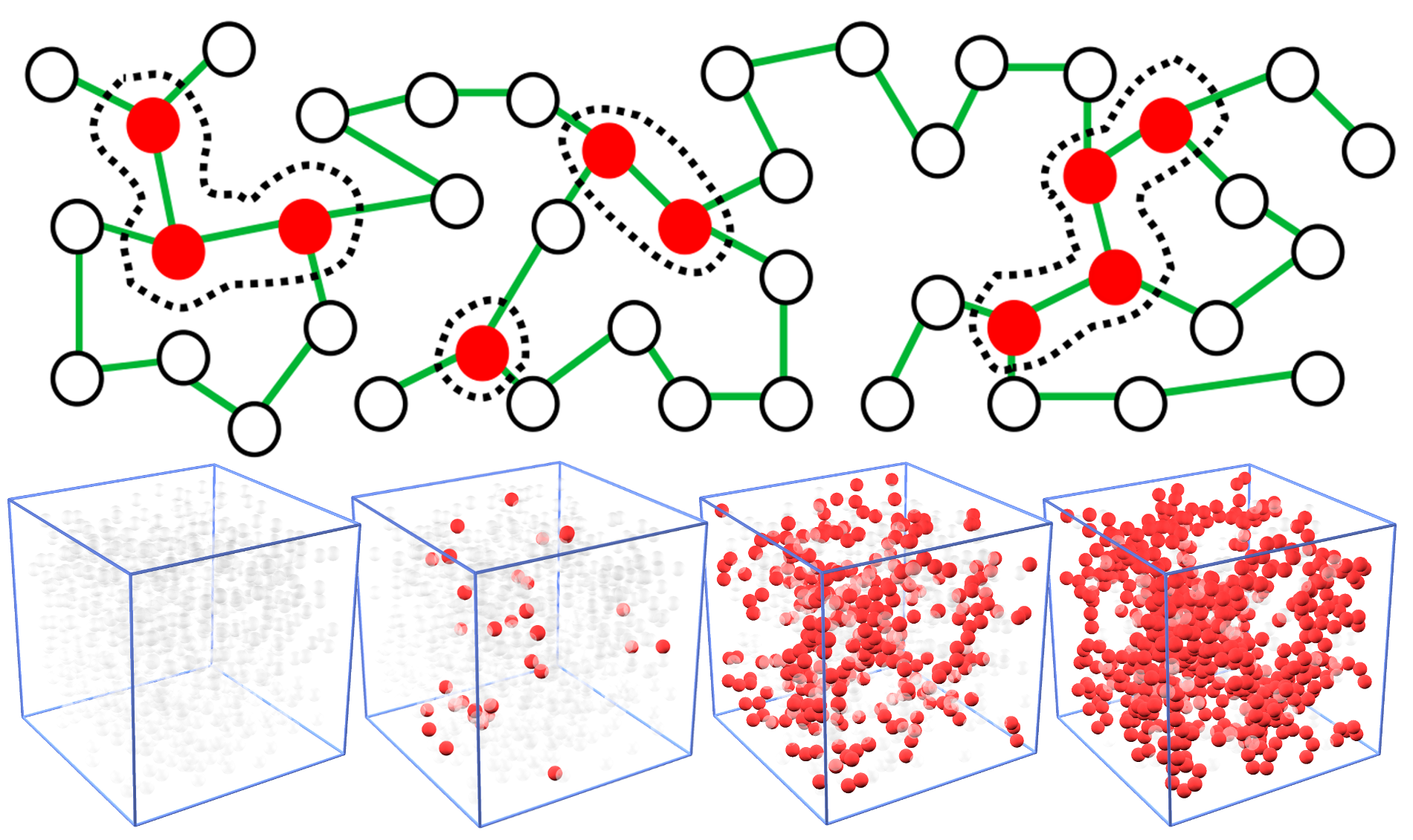} \\
\caption{Relaxation mechanisms responsible for the scale-free dynamics within
the gas-liquid phase coexistence. (top) Schematic representation of the 
relaxation mechanism for network fluids mediated by concerted rearrangements of interconnected particles. 
Red circles represent the liquid phase (three bonds) and black circles represent the gas phase. 
Green lines are the network fluid bonds.
(bottom) Langevin Dynamics of a network fluid quenched from a high 
temperature to a low temperature into the coexistence region. The time increases from left to right for a
system of linear size $L=16$. Red particles represent the liquid phase.} \label{fig.schematic}
   \end{center}
\end{figure*}

\section{Results}
We consider a three-dimensional system of spherical particles of diameter
$\sigma$ with three patches along the equator with an opening angle of
$2\pi/3$. The core-core interaction is pairwise repulsive, described by a
Yukawa-like potential
$V_Y=\frac{A}{\alpha}\exp{\left(-\alpha\left[r-\sigma\right]\right)}$, where
$A$ is the strength of the potential, $r$ is the distance between the particles
centers, and $\alpha^{-1}$ is the range of the interaction. The short-range
patch-patch attraction is described by an inverse Gaussian potential,
$V_G=-\varepsilon\exp(-r_p^2/\xi^2)$, where $\varepsilon$ is the strength of
the attraction, $\xi=0.1\sigma$ the width of the Gaussian, and $r_p$ is the
patch-patch distance. For simplicity, patch-particle interactions are neglected.
In order to follow the collective dynamics, we integrate the corresponding Langevin
equations for the translational and rotational motions (see 
Methods for further details).  

Figure~\ref{fig.diagram}(a) shows the temperature dependence of the 
bonding fraction, $p_b$, defined as the fraction of patches that are bonded. 
The squares are obtained by extrapolating the time-dependence of $p_b$ 
to its asymptotic value from the numerical data for three different system sizes, up to $L=32$,
which is constant within the error bars (see Supplementary Fig.~S2). The solid
(red) line corresponds to the equilibrium value as predicted by Wertheim's
first-order perturbation theory (details in the Methods section). While at high temperatures the numerical equilibrium 
results overlap, at low temperatures the bonding fraction is
systematically lower than the equilibrium value and does not converge to unity
at zero temperature. This signals the presence of an arrested gel in the
coexistence region (see snapshots in Fig.~\ref{fig.diagram}(a) and the equilibrium 
phase diagram in Fig.~\ref{fig.diagram}(b)). Note that the numerical
data are the asymptotic values, suggesting that the thermodynamic structures
are not accessible. 

Far-from equilibrium the asymptotic structures depend on the dynamics. We have
considered two limiting initial configurations with the same number of
particles: A tree-like gel (single aggregate) and a fully unbonded
suspension. The former was grown by ballistic aggregation and loopless
irreversible bond formation \cite{Dias2013b} and corresponds to relaxing a
structure grown at very low temperature \cite{Sciortino2009}. The latter is equivalent to
quenching a colloidal solution from a much higher temperature. Figure~\ref{fig.valence}(a)
shows the time evolution of the fraction of particles with one ($n_1$), two
($n_2$), and three bonds ($n_3$) for both initial configurations at
$k_BT/\varepsilon=0.025$. When starting from a tree-like gel, $n_3$ increases
monotonically, while $n_1$ and $n_2$ decrease, suggesting that structural
relaxation is driven by maximizing the bonded
patches. When starting from a completely unbonded system, the kinetic
pathway is significantly different, resembling a typical aggregation process.
$n_1$ and $n_2$ initially increase due to the formation of small, chain-like,
aggregates but eventually peaks at an intermediate time, when branching and
loop closing dominate. Note that, for the two drastically different initial configurations and pathways, the distributions 
of particles with one, two, and three bonds, are surprisingly similar, sufficiently long times, albeit different from the 
thermodynamic (equilibrium) ones (fully bonded, $n_3=1$). This result suggests the existence of high energy barriers to relax to 
the equilibrium configuration and reveals that the geometrical kinetic structures are statistically robust.

To characterize the dynamics, in the low temperature limit, we follow the convergence of $p_b$ to its asymptotic value $p_b(\infty)$ 
(see Fig.~\ref{fig.valence}(b)).  In Fig.\ref{fig.valence}(b), we observe the size dependence of $p_b(\infty)$ for systems 
ranging from $L=8$ to $L=16$. In this regime, within the coexistence region (see Fig.~\ref{fig.diagram}(b)), 
$p_b$ follows a power law in time, with an exponent of $0.6\pm0.1$. 
The finite-size study shows that the power law regime increases with the size of the system. The
numerical data is consistent with the same exponent for different initial
conditions (tree-like gel and fully unbonded) and temperatures in the
coexistence region (see Supplementary Fig.~S1). 

\begin{figure}[t]
   \begin{center} \includegraphics[width=0.9\columnwidth]{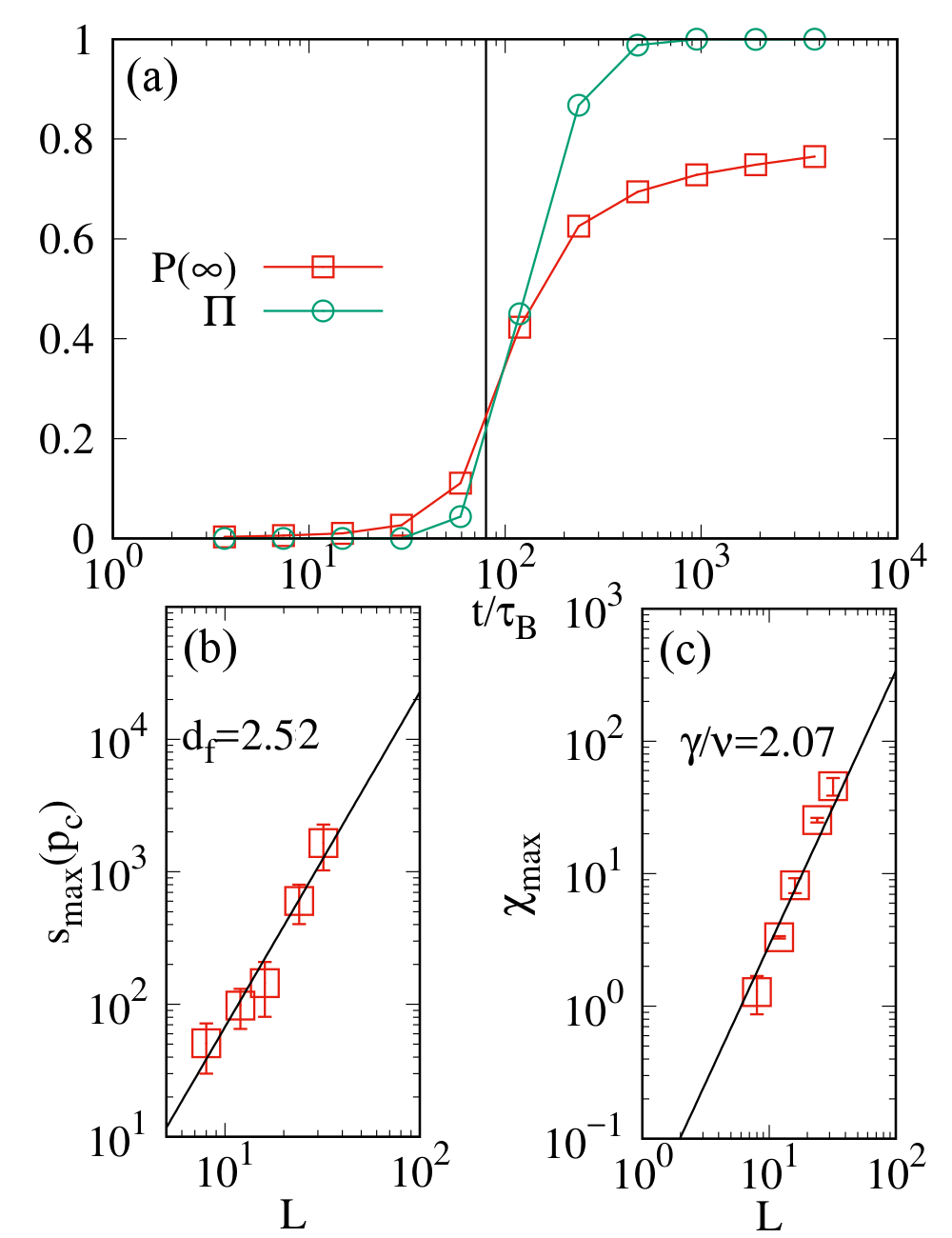} \\
\caption{(a) The fraction of particles in the largest cluster $P(\infty)$ and 
wrapping probability $\Pi$ of the aggregate of liquid particles (with three-bonds) as 
a function of time for a system of 
linear size $L=16$, at $k_BT/\varepsilon=0.025$. Time is
rescaled by the Brownian time $\tau_B=\sigma^2/D_t$, where $\sigma$ is the
particle diameter and $D_t$ its diffusion coefficient.
(b) Size of the largest cluster at the percolation threshold, represented by a vertical solid
line in (a), for system sizes  $L=\{8,12,16,24,32\}$. The solid line represents the theoretical slope
for random percolation where $s_{max}\sim L^{d_f}$ with $d_f=2.52$ the largest aggregate fractal 
dimension at threshold.
(c) Order parameter variance $\chi$, at the percolation threshold, 
for system sizes $L=\{8,12,16,24,32\}$. The solid line represents the theoretical slope
for random percolation where $\chi_{max}\sim L^{\gamma/\nu}$ with $\gamma/\nu=2.07$.
All simulations were performed at a density $0.2\sigma^{-3}$.} \label{fig.percolation}
   \end{center}
\end{figure}

Previous studies for isotropic particles suggest that, within the gas-liquid coexistence region, the 
relaxation dynamics is driven by the formation of individual clusters that coarsen to minimize the 
surface tension \cite{Bray2003}. By contrast here, due to the strong particle-particle interaction, a 
single cluster spanning the entire system is formed well before the scale-free behavior is observed 
(see Supplementary Fig.~S3). This is similar to what has been observed previously in
 other network fluids, namely Laponite \cite{Ruzicka2011b}. There, the heterogeneity of the 
observed cluster suggests a coarsening process, where highly bonded dense 
zones of the aggregate represent the liquid phase. In order to discard that possibility, we assume
that only particles with three bonds are in the liquid phase and identify the clusters of bonded 
particles. We proceed to analyze their relaxation dynamics 
(see a schematic representation in Fig.~\ref{fig.schematic}). Since all particles belong to the gel-like 
network the growth of the liquid phase evolves through the formation of new bonds of particles with 
one or two
bonds. Snapshots of the system at the bottom of Fig.~\ref{fig.schematic} illustrate this behavior. For 
surface tension driven relaxation (coarsening) the characteristic size of the 
clusters $l^*$ should scale as $l^*\sim t^{1/3}$. However, we observe a clearly different exponent,
as $<s>\sim t^{0.3\pm0.1}$  implies $l^*\sim t^{0.10\pm0.04}$ (considering $l\sim <s>^{1/3}$), 
where $<s>$ is the mean cluster size, and suggests a different mechanism
(see Supplementary Fig.~S3 for more details). We should note that the surface tension
of network fluids is ultra-low \cite{Bernardino2012} and this is likely to render coarsening in
these systems rather ineffective.

In Fig.~\ref{fig.percolation} we report the analysis of the kinetics of 
the clusters of liquid particles. Figure~\ref{fig.percolation}(a) shows
the time dependence of the fraction of particles in the largest cluster
and the wrapping probability, defined as the probability of finding a cluster that 
touches opposite sides of the box in the vertical direction. We find a 
percolation transition at $t/\tau_\mathrm{B}\approx 10^2$. Figures~\ref{fig.percolation}(b) and (c)
show the system-size dependence of the number of particles in the largest cluster and its variance 
at the percolation threshold. 
The results are consistent with power-law scaling with the exponents of the 
Random Percolation universality class \cite{Xu2014,Stauffer1994}, suggesting that the relaxation 
dynamics of the liquid clusters goes through a percolation transition. This could explain why the
late time relaxation dynamics of the liquid is independent of the initial condition.  

In summary, we investigated the relaxation dynamics of particles with limited valence, and found a new dynamical 
regime within the coexistence region. The slow thermal reversibility of the bonds induces structural
disorder that is quenched on the experimental timescale effectively hindering
the relaxation of the structure towards thermodynamic equilibrium.
The numerical data for gel aging within the coexistence region is consistent with a
universal exponent, $0.6\pm0.1(\approx2/3)$, for the evolution of the bonding
fraction, which is significantly lower than what is known for surface tension driven coarsening.

\section{Discussion}
Our study reveals a new relaxation mechanism of the gel network, where the liquid does not 
coarsen by interfacial fluctuations as a result of very long lived bonds (practically irreversible)
and ultra-low interfacial tension.
The relaxation mechanism proceeds through the relaxation of the network and not through
local rearrangements of a small number of neighboring particles. 
The liquid phase does not grow by the addition of clusters of nearby particles but through 
geometrical relaxation events occurring over the whole network. At least initially these 
geometrical relaxation events are random. The power-law relaxation is also a hint of 
these collective rearrangements.

Previous studies for Lennard-Jones particles, within the solid-gas coexistence, revealed a power law time dependence 
for the size of the largest cluster at very low temperatures, distinct from (surface tension driven) coarsening \cite{Midya2017}. 
Their numerical data suggests that the exponent 
decreases with the temperature. Although particle diffusion is also strongly suppressed by the gel-like structure, 
our numerical data is consistent with a robust exponent over the range of temperatures considered. 
A crucial difference from the Lennard-Jones fluid is that in the network fluid the liquid clusters are also immobile.

Finally, a comment on other low-valence systems is in order. 
Numerical studies of network fluids of particles with (two) oppositely oriented dipole
moments also suggest a power-law decay of the bond correlation
function~\cite{Schmidle2013,Klapp2016}. These results, obtained by Monte
Carlo (equilibrium) simulations hint at a dynamic slowing down 
driven by gel formation. It would be interesting 
to investigate the dynamics of these dipolar systems and to compare their aging 
to that reported here.
There is also the question of the dependence of the dynamics on the particles valence. 
At low valence, crowding effects are absent. However, the network density increases 
with valence~\cite{Kraft2011,Sabapathy2015,Kern2003,Dias2015} and as the valence
increases, crowding effects are expected to play a role.
Khalil~\textit{et al.} investigated structural arrest transitions in polymer
suspensions with cross linkers~\cite{Khalil2014}, a problem that may be related to mixtures of
limited valence particles. An interplay between the glass and the gel transitions
was observed depending on the polymer and linker concentrations. Future work 
could address similar questions in network fluids and/or their mixtures in order to
elucidate the role of the valence.

\section{Methods}
\textbf{Model and simulations.} We consider spherical particles, all of the same size, with three identical patches distributed along the equator.
The core-core interaction is repulsive, described by a
Yukawa-like potential,
   \begin{equation}
    V_Y(r)=\frac{A}{\alpha}\exp{\left(-\alpha\left[r-\sigma\right]\right)}, \label{eq.Yukawa}
   \end{equation}
where $\sigma$ is the diameter of the particles, $A/(\alpha k_BT)=0.25$ is the
interaction strength and $\alpha/\sigma=0.4$ the inverse screening length. The core-core interaction 
is truncated at a cutoff distance $r_c=1.5\sigma$ (at $r=r_c$ the potential is $10^{-9}A/\sigma$). 
The patch-patch interaction is described by an attractive inverted Gaussian potential \cite{Dias2016,Vasilyev2013}, 
   \begin{equation}
    V_G(r_p)=-\varepsilon\exp(-r_p^2/\xi^2),
   \end{equation}
where $\varepsilon$ is the strength of the attraction, $\xi=0.1\sigma$ the width of the Gaussian, 
and $r_p$ the patch-patch distance. This interaction is truncated also at a cutoff distance $r_{pc}=\sigma$.
For this set of parameters, we expect at most one bond per patch. 

To resolve the stochastic trajectories of the particles, we integrate the Langevin 
equations of motion for the translational degrees of freedom,
\begin{equation}
 m\dot{\vec{v}}(t)=-\nabla_{\vec{r}} V(\vec{r})-\frac{m}{\tau_t}\vec{v}(t)+\sqrt{\frac{2mk_BT}{\tau_t}}\vec{\xi}(t), \label{eq.trans_Langevin_dynamics}
\end{equation}
and the rotational ones,
\begin{equation}
 I\dot{\vec{\omega}}(t)=-\nabla_{\vec{\theta}} V(\vec{\theta})-\frac{I}{\tau_r}\vec{\omega}(t)+\sqrt{\frac{2Ik_BT}{\tau_r}}\vec{\xi}(t).\label{eq.rot_Langevin_dynamics}
\end{equation}
We use the velocity
Verlet scheme with a time step $\Delta t=1\times 10^{-5}$, in units of the Brownian time (time to diffuse one 
particle diameter), and the Large-scale Atomic/Molecular Massively Parallel Simulator (LAMMPS) 
for efficient simulations \cite{Plimpton1995}.
$\vec{v}$ and $\vec{\omega}$ 
are the translational and angular velocities, $m$ and $I$ are the
mass and moment of inertia of the particles, $V$ is the pairwise potential, and $\vec{\xi}(t)$ is the stochastic term 
drawn from a random distribution with zero mean.
We consider the damping time for the translational motion,
\begin{equation}
 \tau_t=\frac{m}{6\pi\eta R}. \label{eq.damping}
\end{equation}
From the Stokes-Einstein-Debye
relation \cite{Mazza2007},
\begin{equation}
 \frac{D_r}{D_t}=\frac{3}{4R^2}, \label{eq.coeff_rel}
\end{equation}
and thus the rotational damping time is $\tau_r=10\tau_t/3$.

\textbf{Wertheim's theory.} The thermodynamic equilibrium properties of the model were calculated using Wertheim's  first order perturbation theory 
(see, e.g.~\cite{Sciortino2007}). The Helmholtz free energy  per particle $f$ is the sum of a reference free energy $f_{ref}$ and a 
perturbation (arising from bonding) $f_b$, $\beta f = \beta f_{ref}+\beta f_b$ where $\beta=1/k_BT$. The reference potential was taken to be 
the repulsive Yukawa, $V_Y(r)$, and the perturbation (that promotes bonding) the attractive patch-patch interaction, $V_G(r_p)$.
The bonding term is given by,
\begin{equation}
\label{fb}
\beta f_b=n\ln (1-p_b)-\frac{n}{2}(1-p_b),
\end{equation}
where $n$ is the number of (identical) patches ($n=3$) and $p_b$ is the bonding fraction, which in the thermodynamic limit 
corresponds to the fraction of bonded patches. We approximate $f_{ref}$ by the hard sphere free energy (in the Carnahan-Starling 
approximation) of a system of particles with an effective diameter $D$ calculated within the Barker-Henderson approximation \cite{Barker1967},
\begin{equation}
\label{DBH}
D=\int_0^{+\infty}\left(1-\exp\left[-\beta V_Y(r)\right]\right)dr.
\end{equation}
The probability $p_b$ was calculated using the law of mass action,
\begin{equation}
\label{lma}
p_b=n\rho\Delta(1-p_b)^2,
\end{equation}
where $\rho$ is the number density defined as the number of particles per unit volume.
$\Delta$ is the integral of the Mayer function of the patch-patch interaction,
\begin{equation}
\label{Delta}
\Delta=\frac{1}{(4\pi)^2}\int d\vec r \int d\hat r_1 \int d\hat r_2 \left[\exp\left(-\beta V_G(r_p)\right]-1\right) g_{ref}(r),
\end{equation} 
where $\vec r$ is the vector between the centers of particles 1 and 2, and $\hat r_i$ is the unit vector that defines the position of a patch on particle $i$ relative to the center of that particle. $g_{ref}(r)$ is the pair correlation function of the reference system, which is approximated by the contact value of the pair correlation function of a system of hard spheres of diameter $D$: $g_{ref}(r)=g_{HS}(r=D)=(1-\eta/2)/(1-\eta)^3$, 
with $\eta=\pi/6D^3\rho$.
The multiple integral (\ref{Delta}) is then reduced to a sum of simple integrals through the following steps (a simple extension of Ref.~\cite{Wertheim1986a}):
\begin{itemize}
\item [1] {Define $z=r_p\equiv |\vec r +\sigma/2(\hat r_2-\hat r_1)|$ and $x=|\vec r + \sigma/2\hat r_2|$.}
\item [2] {Simplify (\ref{Delta}) using  the change of variables:
(i) from $\theta_1$ to $z^2$, in  the integration over $\hat r_1$, with $z^2=
x^2+\sigma^2/4-x\sigma\cos\theta_1$; (ii) from  $\theta_2$ to $x^2$, in the integration over $\hat r_2$, with
$x^2=r^2+\sigma^2/4-r\sigma\cos\theta_2$. One obtains,
\begin{widetext}
\begin{equation}
\label{Deltachv}
\Delta=\frac{\pi g_{HS}(\eta)}{\sigma^2}\int_D^{+\infty}r dr \int_{(r-\sigma/2)^2}^{(r+\sigma/2)^2} \frac{dx^2}{x}
\int_{(x-\sigma/2)^2}^{(x+\sigma/2)^2} f(z^2/\xi^2) dz^2,
\end{equation}
\end{widetext}
where $f(z^2/\xi^2)=\exp\left[-\beta V_G(z)\right]-1$.
}
\item[3] {Change the order of integration in (\ref{Deltachv}), introducing a cut-off $z<\sigma$ (consistent with the procedure adopted in the 
simulations to define a bond) and distinguish the cases $D\ge \sigma$ and $D<\sigma$. After a simple but long calculation, $\Delta$ may 
be expressed as a sum of simple integrals. For $D\ge \sigma$,
\begin{equation}
\label{DeltaDge0}
\Delta=4\pi \sigma^3 g_{HS}(\eta)\alpha^2\int_{-\frac{\delta}{\alpha}}^{\frac{1}{\alpha}} y f(y) F_1(y,\alpha,\delta) dy,
\end{equation}
where $\alpha=\xi/\sigma$, $\delta=(\sigma-D)/\sigma$, and
\begin{equation}
\label{ADelta}
F_1(y,\alpha,\delta)=\frac{\alpha^3 y^3}{6}+\frac{\alpha^2 y^2}{2}+\alpha y\delta\left(1-\frac{\delta}{2}\right)+\delta^2\left(\frac{1}{2}-\frac{\delta}{3}\right).
\end{equation}
For $D<\sigma$,
\begin{equation}
\label{DeltaDlt0}
\Delta=4\pi \sigma^3 g_{HS}(\eta)\alpha^2\left(\Delta_1+\Delta_2+\Delta_3\right),
\end{equation}
where,
\begin{equation}
\label{Delta1}
\Delta_1=\int_0^{\frac{1}{\alpha}}y f(y) F_1(y,\alpha,0)dy,
\end{equation}
\begin{equation}
\label{Delta2}
\Delta_2=\frac{\alpha\delta(2-\delta)}{2}\int_0^{\frac{1-\delta}{\alpha}}y^2 f(y) dy +
\int_{\frac{1-\delta}{\alpha}}^\frac{1}{\alpha} y f(y) F_2(y,\alpha,\delta) dy,
\end{equation}
with,
\begin{equation}
\label{BDelta2}
F_2(y,\alpha,\delta)=\frac{\alpha y}{2}-\frac{\alpha^3y^3}{6}-\frac{(1-\delta)^3}{3},
\end{equation}
and,
\begin{widetext}
\begin{equation}
\Delta_3=\frac{\alpha}{2}\int_0^\frac{\delta}{\alpha}y^2f(y) F_3(y,\alpha,\delta)dy+
\frac{\delta^2}{2}\left(1-\frac{2\delta}{3}\right)\int_\frac{\delta}{\alpha}^{\frac{1-\delta}{\alpha}}yf(y)dy 
+
\int_{\frac{1-\delta}{\alpha}}^\frac{1}{\alpha}y f(y)F_4(y,\alpha,\delta)dy,
\label{Delta3}
\end{equation}
\end{widetext}
with,
\begin{equation}
\label{CDelta3}
F_3(y,\alpha,\delta)=1-\alpha y -(1-\delta)^2,
\end{equation}
and,
\begin{equation}
\label{EDelta3}
F_4(y,\alpha,\delta)=\frac{1}{6}(1-\alpha^3y^3)+\frac{1}{2}(1-\alpha y)(1-\delta)^2.
\end{equation}
}
\end{itemize}

 The percolation line in the $\rho,T$ diagram (see the main text) is obtained by setting $p_b=\frac{1}{n-1}=\frac{1}{2}$ in Eq.~(\ref{lma}). 
 The coexistence line is 
calculated from the equality of the pressures and chemical potentials obtained from $\beta f$. The bonding fraction $p_b$  %(figures pb vs $\rho$ at fixed T and pb vs T at fixed $\rho$) 
is obtained by solving Eq.~(\ref{lma}), in single phase thermodynamic states. 
For state points in the coexistence region ($k_BT/\epsilon<0.0678$ in Fig.~1 at $\rho=0.2$)  $p_b$ is calculated 
using the so-called lever rule: $p_b(\rho,T)=p_{b,l}+x(p_{b,v}-p_{b,l})$, with $x=(\rho_l-\rho)/(\rho_l-\rho_v)$, where $\rho_v$ and $\rho_l$ are 
the coexistence gas and liquid densities at $T$, and $p_{b,v}$ and $p_{b,l}$ are the bonding fractions in the coexisting gas and liquid phases at $T$. 

\begin{acknowledgments}

\textbf{Acknowledgments.} We acknowledge fruitful discussions with Emanuela Del Gado and financial
support from the Portuguese Foundation for Science and Technology (FCT) under
Contracts nos. EXCL/FIS-NAN/0083/2012, UID/FIS/00618/2013, and IF/00255/2013.
\end{acknowledgments}

\bibliography{paper}

\clearpage
%\pagebreak[4]
%\newpage



\section*{Supplementary material}

# Supplementary Information: Relaxation dynamics of a network fluid inside the liquid-gas coexistence region

C. S. Dias,[1,2,*] J. M. Tavares,[2,3,†] N. A. M. Araújo,[1,2,‡] and M. M. Telo da Gama[1,2,§]

[1]*Departamento de Física, Faculdade de Ciências,*
*Universidade de Lisboa, 1749-016 Lisboa, Portugal*
[2]*Centro de Física Teórica e Computacional, Universidade de Lisboa, 1749-016 Lisboa, Portugal*
[3]*Instituto Superior de Engenharia de Lisboa, ISEL,*
*Avenida Conselheiro Emídio Navarro, 1 1950-062 Lisboa - Portugal*

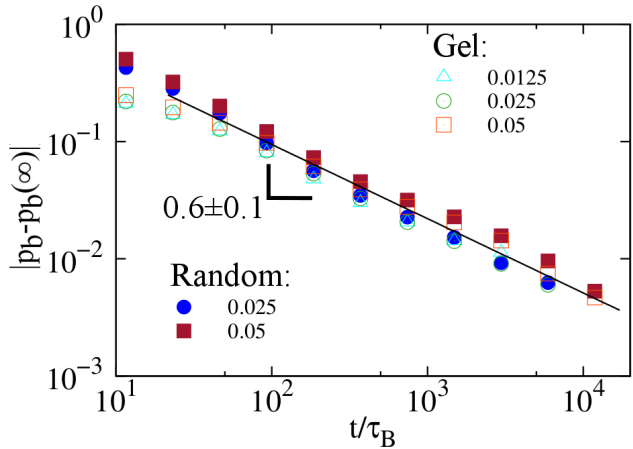

FIG. S1. Convergence of the bonding fraction to the asymptotic value, $|p_b - p_b(\infty)|$, at temperatures $k_BT/\varepsilon = \{0.0125, 0.025, 0.05\}$ starting from a singly connected tree-like gel at temperatures $k_BT/\varepsilon = \{0.025, 0.05\}$ and starting from a fully disconnected suspension, where the time is in units of the Brownian time $\tau_B = \sigma^2/D_t$, $\sigma$ is the particle diameter and $D_t$ the diffusion coefficient. Numerical results were obtained by averaging over 10 samples in a box of length $L = 16$, in units of particle diameter.

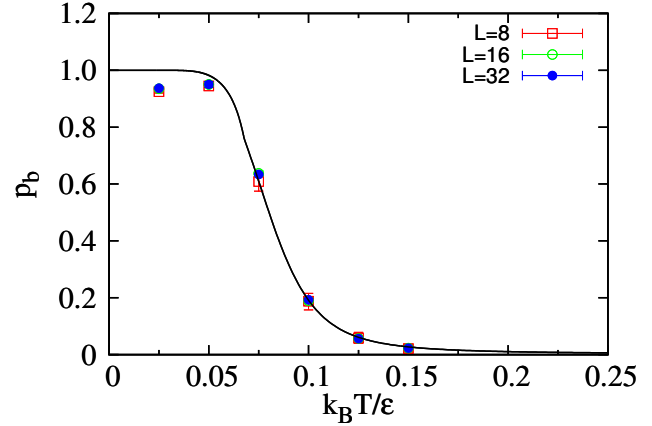

FIG. S2. Bonding fraction $p_b$ as a function of the rescaled temperature $k_BT/\varepsilon$, obtained from the asymptotic extrapolation of the numerical data for systems of linear size $L = \{8, 16, 32\}$. The solid line is the equilibrium value predicted by Wertheim's first-order perturbation theory. Numerical results were obtained by averaging over 10 samples.

## BONDING FRACTION

Figure S1 reveals that the relaxation proceeds as a power law in time, with exponent $0.6 \pm 0.1$, which is observed for different initial conditions and temperatures below $k_BT/\varepsilon = 0.075$.

In Fig. S2 we plot the asymptotic values of $p_b$ for different system sizes showing that size effects are not significant.

## RELAXATION DYNAMICS

In Fig. S3, we plot the average size of clusters as a function of time for all particles and for particles with three bond (assumed to be in the liquid phase). We find that a cluster with all the particles is formed on a timescale much shorter than that corresponding to the scaling regime reported above.

* csdias@fc.ul.pt
† jmtavares@fc.ul.pt
‡ nmaraujo@fc.ul.pt
§ mmgama@fc.ul.pt

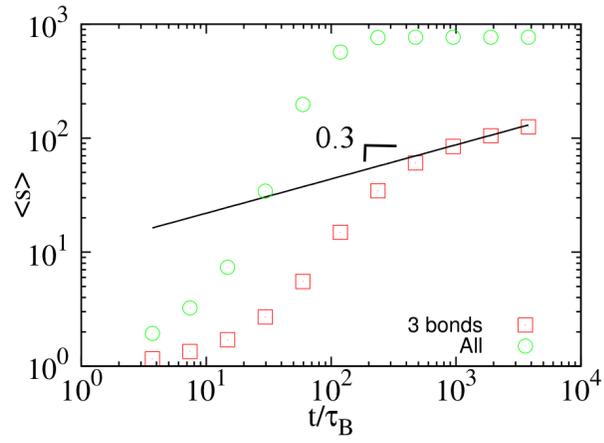

FIG. S3. Average size of clusters $<s>$ of particles (open circles) and of fully bonded particles (open squares) as a function of time. Numerical results obtained by averaging over 10 samples of a system of size $L = 16$.



\end{document}